\documentclass[aps,pra,twocolumn,showkeys,nofootinbib,amsmath,amssymb]{revtex4}
\usepackage{epsf}
\usepackage{color}
\usepackage[rgb,dvipsnames]{xcolor}
\usepackage{dcolumn}
\usepackage{bm}
\everymath{\displaystyle}

\begin{document} 

\title{Exact 
quantum-mechanical equations for particle beams}

\author{A. J. Silenko}
\affiliation{Bogoliubov Laboratory of Theoretical Physics, Joint Institute for Nuclear Research,
Dubna 141980, Russia}
\affiliation{Institute of Modern Physics, Chinese Academy of
Sciences, Lanzhou 730000, China}
\affiliation{Research Institute for
Nuclear Problems, Belarusian State University, Minsk 220030, Belarus}

\begin{abstract} 
The exact quantum-mechanical equations for beams of free particles including photons and for beams of Dirac and spin-1 particles in a uniform magnetic field have been derived. These equations present the exact generalizations of the well-known paraxial equations in optics (exact Helmholtz equation for light beams) and particle physics and of the previously obtained paraxial equation for a 
Dirac particle in a uniform magnetic field. Some basic properties of exact wave eigenfunctions of particle beams have been determined.
\end{abstract}


\maketitle

The paraxial approximation consists in the assumption that the momentum component along the beam direction ($z$ axis) is much larger than the transversal components, $p_z\gg |\bm p_\bot|$. The paraxial equation based on the paraxial approximation
is widely used in optics for studying twisted and untwisted structured light beams \cite{Siegman,Allen,TwPhotons2,TwPhotRev3,BBP}. In this case, it takes the form of the paraxial Helmholtz equation. The connection of this approach with traditional approaches of relativistic quantum mechanics (QM) is considered, e.g., in Refs. \cite{Barnett-FWQM,BliokhSOI,photonPRA}. After the prediction \cite{Bliokh2007} and discovery \cite{UTV} of twisted (vortex) electrons, the paraxial equation is commonly used for their description in vacuum. In Ref. \cite{paraxialLandau}, the paraxial equation has been determined for relativistic electrons in a uniform magnetic field.

As a rule, the paraxial approximation and paraxial equations are used for particle beams delocalized along a certain axis and locaized in two orthogonal directions (2D-localized). Any particle beam describes a multiwave state but can be applied to a \emph{single} particle. We note that particle wave packets can define 3D-localized 
states (like a bullet). In the present study, we adopt the terminology applied in Ref. \cite{BliokhSOI} and use the terms ``particle beam'' and ``particle wave packet'' for 2D- and 3D-localized objects, respectively. While the former term can describe a set of non-interacting noncoherent particles, we will hereinafter consider only a single particle or a set of non-interacting coherent particles being in the same quantum state. The use of this term for the single particle is caused by its multiwave nature. For fermions, one should take into account the Pauli principle forbidding an occupation of the same quantum state by two or more particles. For bosons (e.g., photons), many particles can occupy the same quantum state and form a beam described by the same quantum-mechanical and wave equations as the corresponding single-particle beam. It is important to derive exact quantum-mechanical equations similar to approximate paraxial ones but useful for a description of particle beams with any needed accuracy and applicable beyond the paraxial approximation.

The connection between the relativistic QM and the paraxial equation has been analyzed in Ref. \cite{photonPRA}. The Foldy-Wouthuysen (FW) representation \cite{FW} is very convenient for a determination of this connection. In this representation, relativistic QM has the form equivalent to nonrelativistic Schr\"{o}dinger QM. The Hamiltonian
and all operators in the FW representation are even, i.e., block diagonal (diagonal in two spinors). Relations between the operators in this representation are similar to those between the respective
classical quantities. Quantum-mechanical operators
for relativistic particles in external fields has the same form as in the nonrelativistic quantum theory. In particular, the operators of the position and momentum
are equal to $\bm r$ and $\bm p = -i\hbar\nabla$, respectively \cite{FW,Costella,dVFor,JMP,relativisticFW,Reply2019,PRAFW}.

The FW representation is now effectively used for a \emph{relativistic} description of twisted beams. It has been applied not only for relativistic electrons \cite{Manipulating,ResonanceTwistedElectrons,PhysRevLettEQM2019,photonPRA} but also for photons \cite{Barnett-FWQM,photonPRA}. As a rule, twisted beams were described in the paraxial approximation. The relativistic quantum-mechanical approach has brought some new results. As examples, we can note nonzero effective masses and subluminal velocities of twisted and other structured photons \cite{photonPRA}.

In almost all previous publications, the approximate paraxial equation based on the paraxial approximation was used. Exeptions are Refs. \cite{Karlovets3,Karlovets4} where non-paraxial twisted electron beams as the exact solutions of the relativistic wave equations have been obtained and non-paraxial effects have been predicted.
In the present study, we fulfill the \emph{exact} generalization of the paraxial equation and consider main possibilities of its application.

Another appropriate approach developed and used in many papers (see, e.g., Refs. \cite{Bagrov1,Bagrov2}) employs the light-cone variables $t+z,\, t-z$. This approach yields the Schr\"{o}dinger equation in new variables instead of the Dirac one. The corresponding solutions of the obtained equation have been found both for free bosons and fermions and for charged particles in a magnetic field (see, e.g., Ref. \cite{Bagrov1}). In particular, the mentioned procedure have been done for the vortex beams in a magnetic field (see Ref. \cite{Bagrov2}).

The FW representation is equivalent to the Schr\"{o}dinger one (see Ref. \cite{PRAFW} and references therein) and extends the latter representation to the relativistic region.  Therefore, the approach developed in Refs. \cite{Bagrov1,Bagrov2}, the relativistic FW method, and the approach proposed in the present paper should agree. However, our approach is useful not only for Dirac particles but also for particles (and nuclei) with different spins ($s=0$ and $s>1/2$) and, therefore, it can be applied for obtaining fundamentally new results.

We use the system of units $\hbar=1,~c=1$. We include $\hbar$ and $c$
explicitly when this inclusion
clarifies the problem.

It is convenient to start with the Klein-Gordon-like equation for free particles with \emph{arbitrary} spin:
\begin{equation} \left(\frac{\partial^2}{\partial t^2}-\nabla^2+m^2\right)\varphi=0,
\label{eqKlGorlik}
\end{equation}
where the wave function $\varphi$ has $2s+1$ components for massive particles and two components for massless ones. The number of components of this wave function is defined by the number of \emph{independent} spin components. For massless particles, the helicity is equal to
\begin{equation}
h=\frac{\bm s\cdot\bm p}{p}=\pm s,
\label{helicit}
\end{equation} where $s$ is the spin quantum number.

For spin-1/2 particles, Eq. (\ref{eqKlGorlik}) can be simply obtained by squaring the Dirac equation. Squaring the
first-order Dirac-like equation for spin-3/2 particles (Rarita-Schwinger equation \cite{RaritaSchwinger}) and Proca equations for spin-1 particles (see Ref. \cite{YB}) also leads to Eq. (\ref{eqKlGorlik}). For massive \cite{Weinberg} and massless \cite{WeinbergII} particles with arbitrary spin, Eq. (\ref{eqKlGorlik}) has been rigorously obtained by Weinberg. Other approaches for a derivation of this equation can also be used \cite{BLP}. Thus, Eq. (\ref{eqKlGorlik}) is well substantiated. To derive the
relativistic equation in the FW representation for particles with arbitrary spin, one can fulfill the Feshbach-Villars transformation \cite{FV} (or the generalized Feshbach-Villars transformation involving massless particles \cite{TMP2008}) and the subsequent FW one. For stationary states, the resulting equation reads \cite{ZitterbewegungForMassless}
\begin{equation}
{\cal H}_{FW}\Psi_{FW}=E\Psi_{FW},\qquad
{\cal H}_{FW}=\beta\sqrt{m^2+\bm p^2},
\label{FWelectron}
\end{equation} where $\beta$ is the diagonal matrix multiplying the upper and lower spinors (or spinor-like parts of the wave funcion) by 1 and -1, respectively. When the total energy is positive, the lower spinor vanishes, $\Psi_{FW}=\left(\begin{array}{c} \varphi \\ 0 \end{array}\right)$, and can be omitted. This equation define stationary states with the energy $E$ and the wave function $\Psi_{FW}$. 

Equation (\ref{eqKlGorlik}) can be rewritten in the equivalent form
\begin{equation}
\bm p_\bot^2+p_z^2=P^2,\qquad P\equiv\sqrt{E^2-m^2}.
\label{FWpar}
\end{equation} Amazingly, the same substitution, $$\varphi=\exp(ikz)\Psi,\qquad k=\frac P\hbar,$$ as in Ref. \cite{photonPRA} leads to the exact form of the equation for particle beams: 
\begin{equation}
\begin{array}{c}
\left(\nabla^2_\bot+2ik\frac{\partial}{\partial
z}+\frac{\partial^2}{\partial
z^2}\right)\Psi=0,\\
\nabla^2_\bot=
\frac{\partial^2}{\partial r^2}+\frac1r\frac{\partial}{\partial
r}+\frac{1}{r^2}\frac{\partial^2}{\partial\phi^2},
\end{array}
\label{eqp}
\end{equation}
where $\nabla^2_\bot=-\bm p_\bot^2/\hbar^2$. The form (\ref{eqp}) differs from the well-known one by the presence of the additional operator $\partial^2/(\partial
z^2)$. When the condition $p_z\gg |\bm p_\bot|$ is valid, the expectation value of this operator is small. 
To evaluate this expectation value, we can consider the action of the additional operator on the wave function $\Psi$. For any partial plane wave forming a wave beam, $\Psi=\exp(-ikz)\psi_{FW}(r,\phi)\exp(ik_zz)$ and $$\begin{array}{c}\frac{\partial^2\Psi}{\partial
z^2}=\frac{\partial^2}{\partial
z^2}\exp(-ikz)\psi_{FW}(r,\phi)\exp(ik_zz)\\=-(k-k_z)^2\Psi\approx-\frac{\bm p_\bot^4}{4P^2}\Psi.\end{array}$$ The accuracy of the paraxial approximation is of the order of $\bm p_\bot^2$.

Taking into account the additional term can be necessary for nonrelativistic electron beams. This term can be of importance for optics because of a high precision of optical measurements.  

The result obtained allows us to determine the \emph{exact} Helmholtz equation for light beams. It has the same form as Eq. (\ref{eqp}). However, the wave function should be specified. In QM of the photon, this is the electric field strength $\bm E$ \cite{Barnett-FWQM,photonPRA,ZitterbewegungForMassless}. The same wave function is used in optics \cite{Siegman,Pampaloni}. However, the optical approach gives only the approximate paraxial Helmholtz equation despite numerous attempts to use the full 3-dimensional spatial Helmholtz equation $\nabla^2 A+k^2A=0$ ($A$ is the beam amplitude, see Ref. \cite{Ershkov} and references therein) for a more precise description. Wonderfully, the same substitution as before, $A=\exp(ikz)\Psi,\, k=E/\hbar,\, E$ is the photon energy,
leads to Eq. (\ref{eqp}).

In some cases, the exact paraxial equation can be similarly derived for a charged particle in a uniform magnetic field. Such a possibility exists for a Dirac particle ($s=1/2,\,g=2$) and a spin-1 particle with the gyromagnetic ratio $g=2$. The former case is described by the exact FW Hamiltonan \cite{Case,Energy1,JMP,Energy3}
\begin{equation}
\begin{array}{c}
{\cal H}_{FW}=\beta\sqrt{m^2+\bm{\pi}^2-e\bm\Sigma\cdot\bm B},
\end{array}
\label{eq33new}
\end{equation}
where $\bm{\pi}=\bm{p}-e\bm A$ is the kinetic momentum, $\bm B$ 
is the magnetic induction, and $\beta$ and $\bm\Sigma$ are the Dirac matrices.
This Hamiltonian acts on the bispinor $\Psi_{FW}=
\left(\begin{array}{c} \varphi \\ 0 \end{array}\right)$.
In Ref. \cite{paraxialLandau}, the paraxial equation has been determined in the paraxial approximation. The approach presented above allows us to derive the \emph{exact} equation for particle beams: 
\begin{equation}
\begin{array}{c}
\biggl(\nabla^2_\bot-ieB\frac{\partial}{\partial\phi}-\frac{e^2B^2r^2}{4}+2es_zB+2ik\frac{\partial}{\partial
z}\\+\frac{\partial^2}{\partial
z^2}\biggr)\Psi=0,
\end{array}
\label{eqpar}
\end{equation} where $s_z=\pm1/2$ is the 
spin projection ($\bm B=B\bm e_z$).
As compared with the corresponding equation obtained in Ref. \cite{paraxialLandau}, Eq. (\ref{eqpar}) additionally contains the last term in the round brackets.

The relativistic FW Hamiltonian for a pointlike spin-1 particle with $g=2$ is given by \cite{PhysRevDunitexact}
\begin{equation}
{\cal H}_{FW}=\rho_3\sqrt{m^2+\bm\pi^2-2e\bm S\cdot\bm B},
\label{eqtnHFW}
\end{equation} where $\bm S$ is the spin matrix for spin-1 particles. Such particles are the gauge bosons W$^\pm$.

Importantly, the FW Hamiltonians (\ref{eq33new}) and (\ref{eqtnHFW}) remain valid for a nonuniform magnetic field.

The exact equation for beams of spin-1 particles has the form (\ref{eqpar}) but $s_z=-1,0,+1$. It is easy to obtain the corresponding equation for spinless particles (and nuclei).
The use of the exact equation for particle beams instead the approximate paraxial one is important when one considers such Laguerre-Gauss beams in a uniform magnetic field which undergo spatial oscillations \cite{arXiv}.

The form of the additional term in the exact equations (\ref{eqp}) and (\ref{eqpar}) defines some basic properties of exact particle wave eigenfunctions. For Gaussian, Laguerre-Gauss, and Hermite-Gauss beams, this term does not change quantum numbers of eigenstates
\cite{footnote}. As a result, the exact particle wave eigenfunctions conserve the transversal structure of the corresponding eigenfunctions obtained in the paraxial approximation.

In summary, we have derived the exact quantum-mechanical equations for beams of free particles with an \emph{arbitrary} spin
including photons and for beams of particles with $g=2$ and spins 1/2, 1 in a uniform magnetic field. These equations present the exact forms of the well-known paraxial equations in optics (i.e., the exact Helmholtz equation for light beams) and particle physics and of the corresponding equation \cite{photonPRA} for a 
Dirac particle in a uniform magnetic field. Some basic properties of exact wave eigenfunctions for particle beams have been determined. Importantly, our equations being equivalent to the corresponding equations of relativistic QM in the FW representation allow describing particle beams with any needed accuracy and remain applicable beyond the paraxial approximation.

\medskip
The author acknowledges the support by the National Natural Science
Foundation of China (grants No. 11975320 and No. 11805242) and by the Chinese Academy of Sciences President's International Fellowship Initiative (grant No. 2019VMA0019).


\begin{thebibliography}{}
\bibitem{footnote} For Hermite-Gauss beams, the operator $\nabla^2_\bot$ should be written in the form $\nabla^2_\bot=\frac{\partial^2}{\partial x^2}+\frac{\partial^2}{\partial y^2}$.
\end{thebibliography}
\end{document}